# Modulation of skyrmionic magnetic textures in two-dimensional vdW materials and their heterostructures

Xiaoyan Yao[1,*], Di Hu[1], and Shuai Dong[1,†]

**ABSTRACT**

The intrinsic magnetism observed in two-dimensional (2D) van der Waals (vdW) materials provides a unique opportunity for exploring the 2D topological magnetic textures, in particular skyrmionic magnetic textures (SMTs) including skyrmion and its topological equivalents. Since the experimental discovery of skyrmions in the 2D vdW materials and their heterostructures, a critical challenge lies in the control of these SMTs to translate their intriguing features into spintronic applications. Here, we review the recent experimental and theoretical progresses on the modulations of SMTs in 2D vdW monolayer materials and their heterostructures. Besides well-established basic modulation factors including temperature, magnetic field and sample thickness, we present the experimental realization of mobility and transition driven by electric current, and the theoretical prediction of diverse magnetoelectric modulations by electric field. Considering the 2D character of vdW layered materials, strain and stacking style are also efficient approaches to tune the magnetic textures.

[1]School of Physics, Southeast University, Nanjing 211189, P. R. China

*Correspondence: yaoxiaoyan@seu.edu.cn (X.Y.)

†Correspondence: sdong@seu.edu.cn (S.D.)

## INTRODUCTION

Since the single-layer graphene as the first true two-dimensional (2D) material was successfully isolated from graphite in 2004[1], the magnetism has been pursuing in 2D materials. Whereas the Mermin-Wagner theorem states that long-range magnetic order cannot exist in 2D isotropic Heisenberg model at finite temperature due to thermal fluctuations[2], the magnetic anisotropy plays an essential role in overcoming the thermal fluctuations down to a monolayer limit, and thus to realize the long-range magnetic order in 2D materials[3,4]. Besides various efforts to attach magnetism to nonmagnetic 2D materials[5], in 2017, the intrinsic 2D ferromagnetism was first observed in van der Waals (vdW) layered $Cr_2Ge_2Te_6$[6] and $CrI_3$[3], which is particularly intriguing due to a wide range of opportunities for fundamental researches and technological applications[7-9].

The realization of long-range magnetic order in 2D magnets provides a unique opportunity for exploring the 2D topological magnetic textures with nonzero topological charge ($Q$). Among them, skyrmionic magnetic textures (SMTs), including skyrmion and its topological equivalents, such as bimeron with in-plane magnetization, are much highlighted due to their nanoscale dimensions, topologically protected stability and potential applications for nonvolatile energy-efficient spintronic devices[10-14]. As natural excitations of ferromagnet in 2D space, SMTs are much expected to exist in 2D magnetic materials, and the previous experiments demonstrated that the reduction of materials' dimension is beneficial to the stability of skyrmions[15]. In 2019, it was inspiring that skyrmions were observed experimentally in the 2D vdW $Cr_2Ge_2Te_6$[16], $Fe_3GeTe_2$[17] and later in their heterostructures[18-21]. Very recently, a room-temperature skyrmion lattice was also reported in experiment of 50% Co-doped $Fe_5GeTe_2$[22]. Besides the experiments, many efforts have been made on theoretical and computational investigations to predict more 2D vdW SMTs-hosting materials, such as $MnBi_2Te_4$[23], $MnXTe$ (X=S and Se)[24,25], $CrInX_3$ (X = Te, Se)[26], $CrTe_2$[27] and bilayer $Bi_2Se_3$-EuS[28]. In the noncentrosymmetric materials or interfacial symmetry-breaking heterostructures, SMTs are mainly induced by the Dzyaloshinskii-Moriya interaction (DMI) resulting from spin-orbit coupling with inversion asymmetry. While in the centrosymmetric materials, the main origins could be the long-ranged magnetic dipole interactions or the frustration of exchange interactions[29-32]. For the thin magnetic film, the dipole-dipole interaction may dominate the stabilization of skyrmion.

With the developments of 2D materials and topological magnetism, a key challenge for applications resides on the efficient manipulation of topological magnetic textures in 2D materials. In this paper, we review very recent investigation advances on the direct external modulations of SMTs in 2D vdW materials and their heterostructures. The main text is arranged by the representative modulation approaches. Besides well-established basic modulation factors including temperature, magnetic field and sample thickness, we discuss the experimental realization of electric-current-driven mobility and the theoretical prediction of diverse magnetoelectric modulations by electric field. Considering the 2D character of vdW layered materials, strain and stacking style are also efficient approaches to tune the magnetism. Finally, a summary with brief prospect is presented. It is anticipated that this review will shed light on the exploitation of more effective modulations on SMTs, and thus help to the future spintronic applications.

**REPRESENTATIVE MODULATION APPROACHES**
1) Temperature, magnetic field and specimen thickness

The generation of SMTs results from a delicate balance between the various interactions, anisotropy, thermal fluctuation, and external magnetic field. When exchange and DMI interactions satisfy certain conditions, SMTs appear in a certain region of temperature and magnetic field[17,20,22]. Thus, temperature and magnetic field are basic external approaches to control these SMTs, which had been confirmed in experiments repeatedly[21,22]. Besides the creation or annihilation of SMTs, they extensively affect the SMT's size. Usually the size of a single skyrmion decreases with increasing magnetic field[16,17,33], as illustrated in Fig. 1(a). The influence of temperature is more complex. When the temperature rises, the size of a single skyrmion diminishes in most cases[18,20,22], but sometimes it also expands[21]. In addition, the experiments demonstrated that the magnetism may be strongly dependent on the history of magnetic field and temperature[22]. The magnetic field can be tuned to switch different magnetic phases. The calculation predicted magnetic-field-induced topological transitions, such as one from a spontaneous antiskyrmion lattice to a standard skyrmion lattice in monolayer $NiI_2$[34], and another of skyrmion–bimeron–ferromagnet phase transition cycle in $WTe_2/CrCl_3$ heterostructure[35]. On the other hand, it was also observed in quite a few experiments that the skyrmions may remain stable as a metastable state even after removing the external field[17,20-22].

Another important factor considered by most experiments is specimen thickness. It was reported that the skyrmions appear in only a narrow thickness range[22,36]. Moreover, the skyrmions' size can be strongly dependent on the thickness of sample. Usually, the average skyrmion diameter increases with increasing specimen thickness, following Kittel's law (the domain size depends on the square root of sample thickness) determined by minimizing the total energy including magnetostatic energy and domain wall energy[37,38]. For example, it was found that the skyrmion size in $Fe_3GeTe_2$ grows from about 100 to 750 nm as the specimen thickness increases from about 90 nm to 2 μm[33], as illustrated in Fig. 1(b). Furthermore, recent experiment revealed a transition from Bloch-type bubbles to Néel-type skyrmions by modulating sample thickness of 2D vdW $Fe_{5-x}GeTe_2$, as displayed in Fig. 1(c). The calculations demonstrated that the surface modulation plays a key role in introducing DMI to the $Fe_{5-x}GeTe_2$ cleavage surface, and thus promote the generation of Néel-type skyrmions[38].

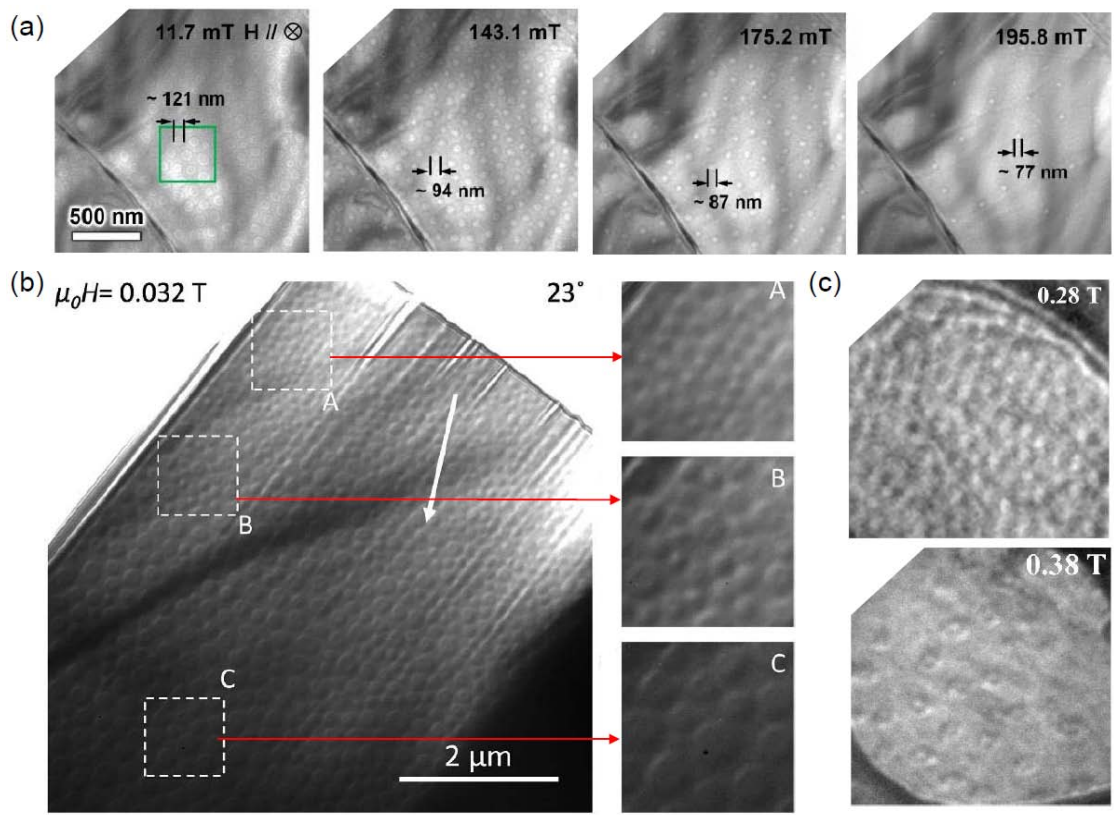

Fig. 1 **SMTs modulated by magnetic field and specimen thickness.** a) The Lorentz transmission electron Microscopy (LTEM) images showing the skyrmionic bubbles under different external magnetic fields in $Cr_2Ge_2Te_6$. Reprinted from Ref. 16. Copyright © 2019, American Chemical Society. b) The skyrmion size dependent on the thickness of $Fe_3GeTe_2$ lamella. LTEM overview of a wedge-shaped lamella (left), where the white arrow presents the direction of increasing thickness and white squares exhibit three regions of different thicknesses (1 μm × 1 μm in area) with magnified views shown on right. Reprinted from Ref. 33. Copyright © 2022, The Authors. c) LTEM images of skyrmions dependent on thickness of $Fe_{5-x}GeTe_2$ flake at 130 K. Top: Néel-type skyrmions in the flake with a thickness of about 61 nm under the magnetic field of 0.28 T. Bottom: Bloch-type bubbles in the flake with a thickness of about 116 nm at 0.38T. Reprinted from Ref. 38. Copyright © 2022, American Physical Society.

2) Electric current

It was much highlighted that in some typical skyrmionic bulk materials, such as MnSi and FeGe, the current density required to move skyrmions is very low, i.e. a few $10^6$ A/m$^2$, which is about five orders of magnitude smaller than those required to move magnetic domain walls (about $10^{11}$ to $10^{12}$ A/m$^2$)[39-42]. It is highly anticipated that such a low driving current density can also be realized in vdW 2D materials. Recently, the experiment on the ferromagnetic Fe$_3$GeTe$_2$-based vdW heterostructures showed that the transition from labyrinth random domain to SMTs can be triggered by strong pulses of in-plane electric current[20], as shown in Fig. 2(a). Moreover, an individual skyrmion can be driven by nanoseconds current pulses, and the propagation direction is along the electron-flow direction, as shown in Fig. 2(b). The average skyrmion velocity was measured to be about 1m/s at a moderate current density of $1.4 \times 10^{11}$ A/m$^2$, below which no motion is observed[20]. Very recently, the room-temperature skyrmion motion driven by electric current was reported experimentally in 50% Co-doped Fe$_5$GeTe$_2$ above the threshold current density of about $10^6$ A/cm$^2$ [22]. The skyrmion drift velocity was estimated to be about 35.4 m/s at a current density of about $4.76 \times 10^6$ A/cm$^2$. From the theoretical aspect, the current-driven skyrmion motion can be understood by the spin torque mechanism, including spin transfer torque and spin orbit torque in different geometries for the injection of spin-polarized current [43]. The calculation on the multiferroic LaCl/In$_2$Se$_3$ heterostructure predicted that an in-plane current of $3 \times 10^{10}$ A/m$^2$ can be applied to mobilize the bimerons at the speed of 72.17m/s[44]. In addition, it was predicted that in a frustrated ferromagnetic monolayer, an isolated bimeron could be driven by current into elliptical motion accompanied by rotation [45].

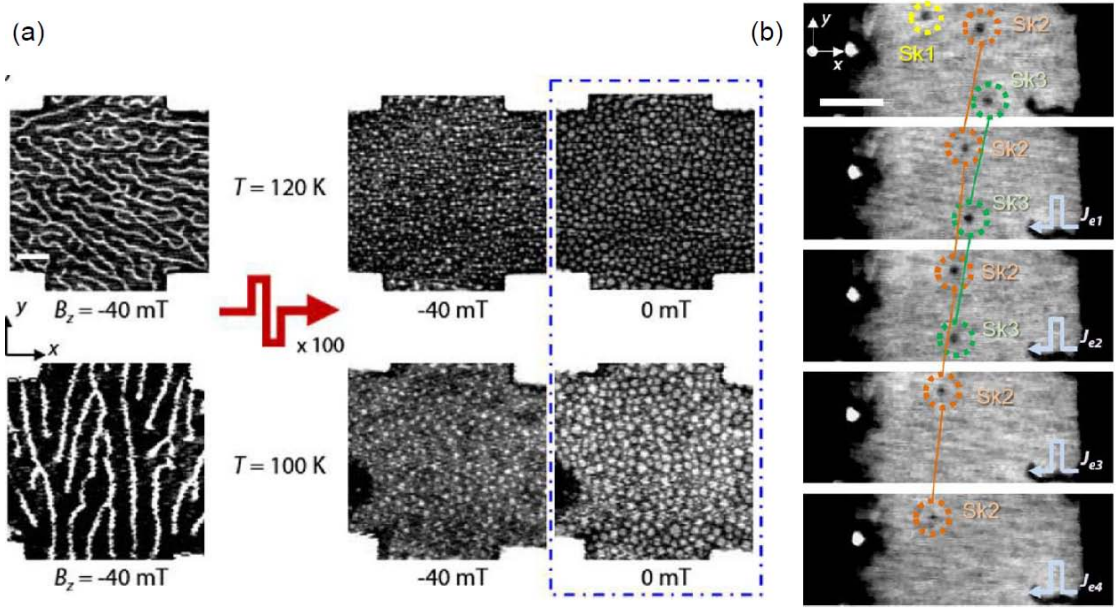

Fig. 2 SMTs modulated by electric current in Fe$_3$GeTe$_2$-based heterostructure. a) The scanning transmission x-ray microscopy (STXM) images. The two images in middle were obtained after the application of bipolar pulse bursts on the left states with initial labyrinth domains at 120 and 100 K respectively, and the other two images on right were obtained after removing magnetic fields. Scale bar: 1 μm. b) Sequential STXM images showing skyrmions

at the oblique field of −20 mT and 100 K, where each image was obtained after injecting five unipolar current pulses along the +x direction with an amplitude of $1.4 \times 10^{11}$ A/m$^2$ and duration 50 ns. Individual skyrmions (Sk1, Sk2, and Sk3) are outlined in colored circles for clarity. Scale bar: 1 μm. Reprinted from Ref. 20. Copyright © 2021, American Physical Society.

3) Electric field

Compared to the electric current, applying an electric field is more energy-efficient to control SMTs, which can be used in much-desired low-power-consumption spintronic devices. The electric-field creation and annihilation of skyrmions had been realized experimentally in the non-vdW epitaxial film[46], multilayers[47,48] and single crystal bulks[49,50]. It was also proposed that an out-of-plane electrical field applied on ferromagnetic CrI$_3$ monolayer induces ionic displacements and DMI, and thus produces sub-10-nm Néel-type magnetic skyrmions[51]. Compared with the direct influence of electric field on exchange couplings, another more feasible approach to electrically control SMTs is to apply switchable electric polarization and magnetoelectric coupling effect. The most achievable way is to construct heterostructure by integrating ferroelectric and magnetic materials, and thus the sizable DMI can be induced by the interfacial symmetry breaking. Since DMI is sensitive to the interface orbital hybridization and charge transfer effect, the electric polarization is expected to effectively control DMI in adjacent magnetic system, and thus to control the magnetic textures[52]. It had been reported experimentally that skyrmions can be efficiently manipulated by electric fields via polarization switching in BaTiO$_3$/SrRuO$_3$ perovskite heterostructure[53].

In contrast to the perovskite-based heterostructures, 2D vdW multiferroic heterostructures are expected to be more advantageous in achieving strong magnetoelectric coupling, since more atoms exposed to the surface make magnetism more sensitive to electric polarization[44]. The recently discovered 2D vdW ferroelectric materials with switchable spontaneous polarization[54,55], such as In$_2$Se$_3$[56-58], just can be used to achieve such an electric control of SMTs in 2D vdW heterostructures. Based on ferroelectric In$_2$Se$_3$ and different ferromagnetic monolayers, various multiferroic vdW heterostructures have been constructed theoretically, exhibiting diverse electric modulation phenomena. In 2020, Wei Sun et al. proposed a LaCl/In$_2$Se$_3$ heterostructure consisting of 2D easy-plane ferromagnetic metal LaCl and In$_2$Se$_3$ layers[44]. The out-of-plane electric polarization from In$_2$Se$_3$ layer breaks the spatial inversion symmetry in the adjacent LaCl layer, giving rise to a DMI. By switching the electric polarization, the anisotropy and exchange interactions were manipulated. Thereby the generation and removal of magnetic bimerons with about 23 nm diameter can be achieved[44], as presented in Fig. 3(a). Similarly, an electric-field-controlled writing and deleting process of skyrmions was predicted in the multiferroic Cr$_2$Ge$_2$Te$_6$/In$_2$Se$_3$ vdW heterostructure[59]. When the direction of electric polarization in $In_2Se_3$ is switched by electric field, the strength of DMI changes so significantly that the magnetic structure of $Cr_2Ge_2Te_6$ will switch between the topologically distinct skyrmion lattice and the ferromagnetic state[59]. The electric control of topological magnetic phases was also predicted in $MnBi_2Se_2Te_2/In_2Se_3$ multiferroic heterostructure[60]. By flipping the direction of electric polarization from up to down, the magnetic anisotropy changes from in-plane to perpendicular type. Thus, bimerons with opposite topological charges ($Q = \pm 1$) can be obtained by applying an in-plane magnetic field in the heterostructure with up polarization, while skyrmions with uniform $Q$ can be obtained by applying an out-of-plane magnetic field in the heterostructure with down polarization[60], as plotted in Fig. 3(b).

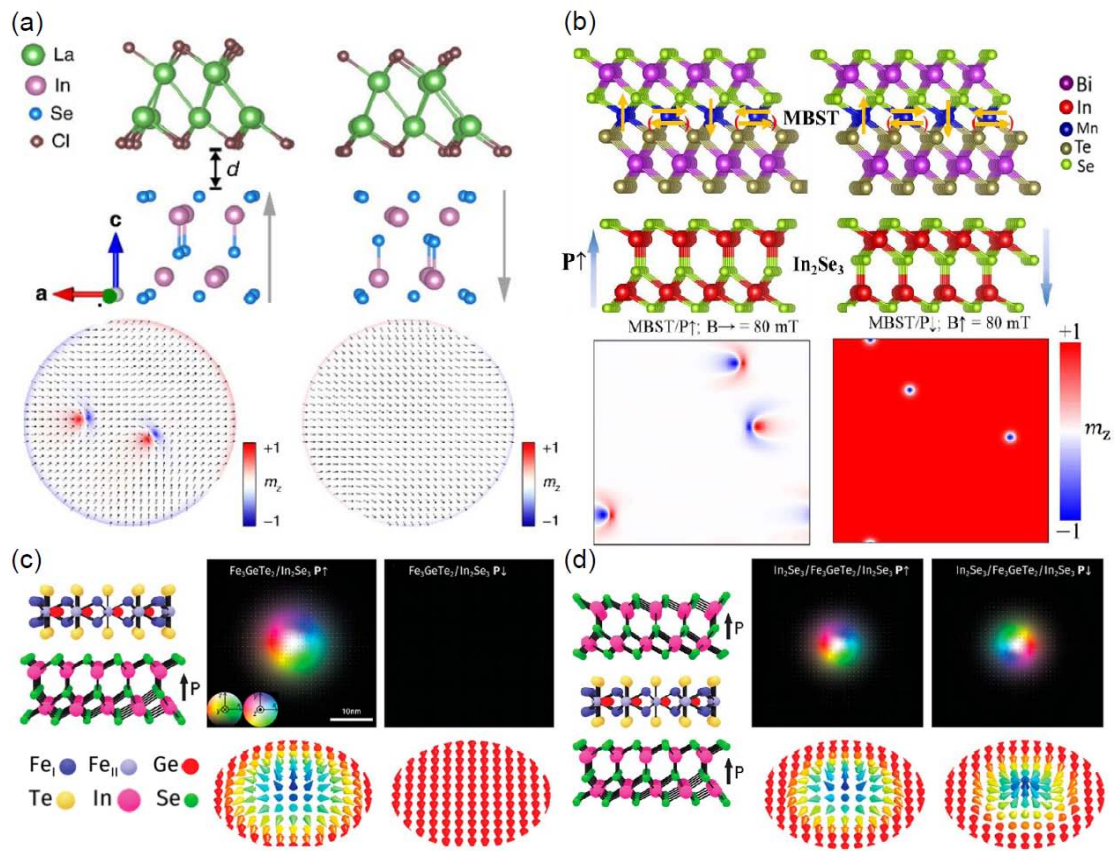

Fig. 3 Ferroelectrically tunable SMTs in 2D vdW multiferroic heterostructures. a) Top: side views of $LaCl/In_2Se_3$ heterostructures with opposite polarizations. Bottom: the corresponding magnetic configurations. Reprinted from Ref. 44. Copyright © The Authors. b) Top: side views of $MnBi_2Se_2Te_2/In_2Se_3$ heterostructures with opposite polarizations. Bottom: the corresponding magnetic configurations with external magnetic fields labeled above each panel. Reprinted from Ref. 60. Copyright © 2021, American Physical Society. c) Left: $Fe_3GeTe_2/In_2Se_3$ vdW heterostructure. Right: the magnetic configurations for opposite polarizations. d) Left: $In_2Se_3/Fe_3GeTe_2/In_2Se_3$ vdW heterostructure. Right: the magnetic configurations for opposite polarizations. Reprinted from Ref. 65. Copyright © 2022,

American Chemical Society. The arrows represent the direction of electric polarization (P) of In$_2$Se$_3$.

Besides the heterostructures with coupling magnetic and ferroelectric layers, the recent discoveries of 2D multiferroic vdW layered materials with spontaneous electric and magnetic polarizations opened a new route towards the highly-desired control of topological magnetism by means of electric field. The inversion symmetry breaking induced by ferroelectric displacement leads to a sizable DMI, which enables the stabilization of SMTs in sub-10 nm size. Meanwhile, the direction of electric polarization determines the chirality of DMI, and thereby determines spin swirling direction and the morphology of SMT. As schematically shown in Fig. 4(a), two stable states with opposite electric polarization (P) and associated DMI show a double-well energy landscape, which can be switched by an out-of-plane electric field[61]. Considering the direction of magnetization in the skyrmion core (m) additionally, four energetically degenerate skyrmionic states with different direction combinations of P and m, can be switched between each other by applying external electric and magnetic fields[61]. This concept was predicted to be realized in various 2D multiferroic monolayers, such as CrN[61], Co(MoS$_2$)$_2$ (bilayer MoS$_2$ with Co intercalated)[62], CuCrP$_2$Se$_6$ and HgInP$_2$O$_6$[63]. Different from skyrmions in first three materials with the perpendicular magnetic anisotropy, HgInP$_2$O$_6$ monolayer with the in-plane anisotropy possesses isolated asymmetrical bimerons. As illustrated in Fig. 4(b), the opposite electric polarizations just correspond to the opposite chirality of DMI[62]. By sweeping an out-of-plane electric field, the sign of DMI parameter d changes synchronously with the electric polarization P switching[63], as illustrated in Fig. 4(c). At the same time, harnessing the electric field also causes the magnitude variations of anisotropy and exchange interactions, and so affects the SMT's radius. Thus, due to the intrinsic magnetoelectric coupling in 2D multiferroics, the effective control of SMTs with tunable size and reversible chirality can be realized by applying electric field to control the electric polarization and DMI.

Besides the electric-field control of SMT's morphology as mentioned above, the topological charge can also be switched by electric field. It was predicted that the multiferroic VOI$_2$ monolayer may host isolated magnetic bimerons as metastable state[64]. As plotted in Fig. 4(d), the displacement of V along the *a*-axis leads to an electric polarization, and simultaneously breaks inversion symmetry, resulting in a significant DMI with only z component $D_z^I$. When the electric field flips the electric polarization, $D_z^I$ changes its magnitude and switches its sign, as shown in Fig. 4(e). By removing and regaining DMI, the electric field changes energy landscape. Meanwhile the magnetic easy axis changes from the in-plane direction near the ferroelectric ground states to the out-of-plane direction near the intermediate paraelectric state, as shown in Fig. 4(e). The out-of-plane anisotropy appearing near the paraelectric state enables the topologically trivial ferromagnetic state with $Q = 0$, bridging nontrivial bimerons with $Q = -1$ and $+1$ in ferroelectric states. All these concomitantly lead to an electric-field switching of bimeron's topological charge between $Q=-1$ and $1$ in a controllable and reversible fashion[64], as displayed in Fig. 4(f).

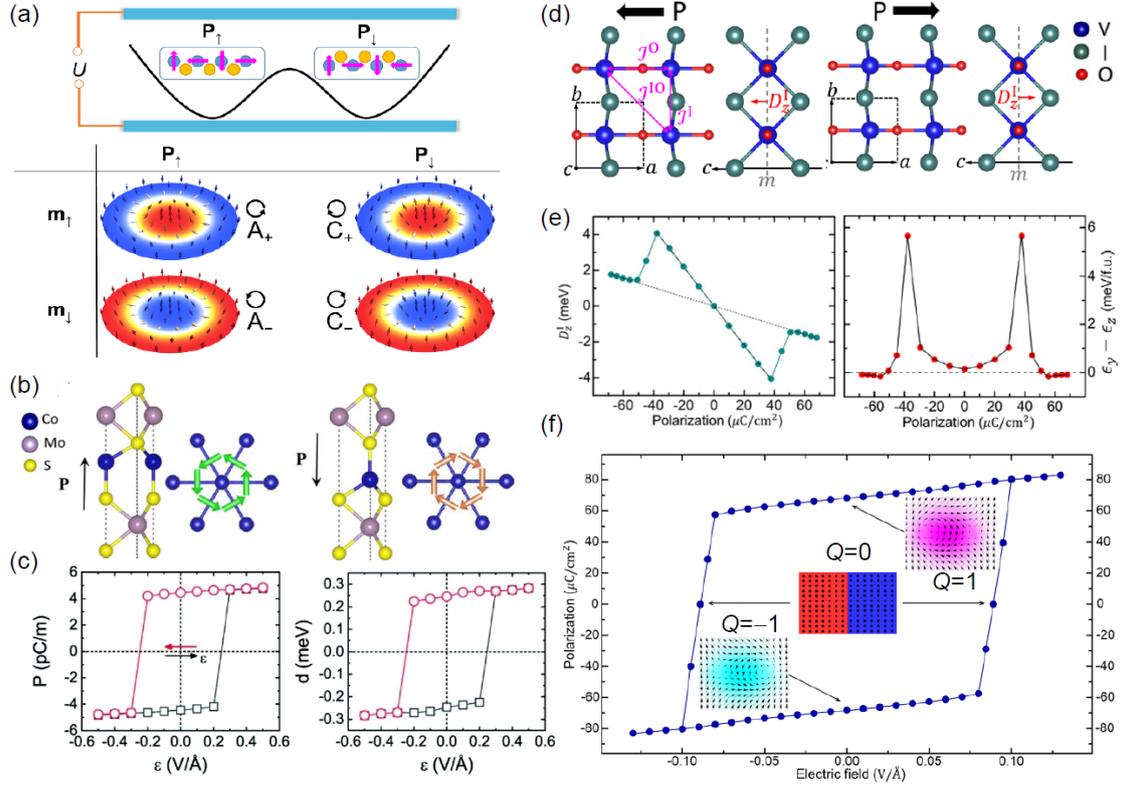

Fig. 4 Ferroelectrically tunable SMTs in 2D vdW multiferroic monolayers. a) Sketch of the switchable multiferroic skyrmions. Top: two stable states in the double-well energy landscape with opposite polarization P and different chirality (magenta arrows), which can be switched by electric field. Bottom: the four interswitchable multiferroic skyrmions with different directions of P and m. Reprinted from Ref. 61. Copyright © 2020, American Physical Society. b) The ferroelectric $Co(MoS_2)_2$ monolayer with opposite P and associated DMI of different chirality (colorful arrows). Reprinted from Ref. 62. Copyright © 2022, American Physical Society. c) Calculated P and DMI parameter d of $CuCrP_2Se_6$ monolayer as the out-of-plane electric field ε sweeps, where black and red points represent different sweep directions. Reprinted from Ref. 63. Copyright © 2021, The Royal Society of Chemistry. d) The ferroelectric $VOI_2$ monolayer with opposite P, where the red arrow presents the direction of DMI. The $a$, $b$ and $c$ lattice vectors are parallel to $x$, $y$, and $z$ directions, respectively. e) The DMI parameter $D_z^I$ and the magnetic anisotropy energy ($\epsilon_y$-$\epsilon_z$) as a function of electric polarization. $\epsilon_y$ and $\epsilon_z$ are the energies of the ferromagnetic states with magnetization lying along y and z directions respectively. f) The polarization-electric field loop, where the insets present the spin patterns with different topological charges. Reprinted from Ref. 64. Copyright © 2020, American Physical Society.

Comparing to the chirality reversal of DMI controlled by electric polarization in multiferroic monolayers, the ferroelectric-controlled asymmetric variation of DMI usually appears in multiferroic heterostructures as mentioned above. Another example is $Fe_3GeTe_2/In_2Se_3$ vdW heterostructure, in which the ferroelectric-controlled DMI variation from a large to a small value was predicted to induce a switchable generation and removal of skyrmion[65]. However, it was interesting that in an $In_2Se_3/Fe_3GeTe_2/In_2Se_3$ vdW heterostructure, the sign of DMI can be switched by reversing electric polarization, which induces the chirality variation of skyrmions similar to the case in multiferroic monolayers[65], as plotted in Fig. 4(d).

4) Strain

Magnetic properties are closely related to the structural parameters of materials, and strain provides a powerful approach for tuning the magnetic parameters by directly influencing the structure, which can be theoretically described by magneto-elastic coupling [66]. For 2D vdW layered materials with flexibility, the strain modulation of magnetic textures will be more efficient. By interfacing with various substrates or fabricating stretchable heterostructures, the tunable strain can be realized. The recent calculation results highlighted the possibility of inducing and tuning SMTs in 2D magnets through strain engineering. The calculation proposed a $Fe_3GeTe_2$/germanene vdW heterostructure as illustrated in Fig. 5(a)[67]. No magnetic skyrmion emerges in this heterostructure without strain. When a small compressive strain (−3%) is applied, the magnetic anisotropy energy (MAE) is dramatically reduced to 25% and the DMI is significantly enhanced up to 400% of its original value (Fig. 5(b)), which stabilizes nanoscale Néel-type skyrmions at zero magnetic fields, as plotted in the inset of Fig. 5(c). Such an efficient strain control of DMI and MAE is general for $Fe_3GeTe_2$ heterostructures with buckled substrates, which also tunes the size of single skyrmion as demonstrated in Fig. 5(c)[67].

Up to now, DMI is still the central reason to generate SMTs. For the vdW 2D magnets, the sizable DMI can be obtained by fabricating the Janus structure with different atoms occupying top and bottom layers to break the inversion symmetry. Some of them were predicted to possess SMTs, and strain may be an effective approach to tune these SMTs, such as the Janus monolayer CrXTe (X = S, Se, as shown in Fig. 5(d))[68] and $Cr_2X_3Y_3$ (X,Y = Cl, Br, I, X≠Y, as shown in Fig. 5(g))[69]. Nanoscale skyrmions can be stabilized by applying external magnetic field in the unstrained CrSeTe and $Cr_2I_3Cl_3$ monolayers. Usually tensile strain enhances the perpendicular magnetic anisotropy obviously, while compressive strain reduces it and even induces a switch to in-plane anisotropy in CrSeTe and $Cr_2I_3Cl_3$, as presented in Fig. 5(e) and 5(h). The strain effect on DMI is more complex, i.e. different tendencies emerge for CrXTe and $Cr_2X_3Y_3$, as displayed in Fig. 5(f) and (i). In particular, a sign reversal of DMI can be controlled by strain in CrSTe and $Cr_2Cl_3Br_3$. Together with the strain-modulation on complicated exchange couplings, these results provide conditions to tune both size and morphology of SMTs by controlling strain.

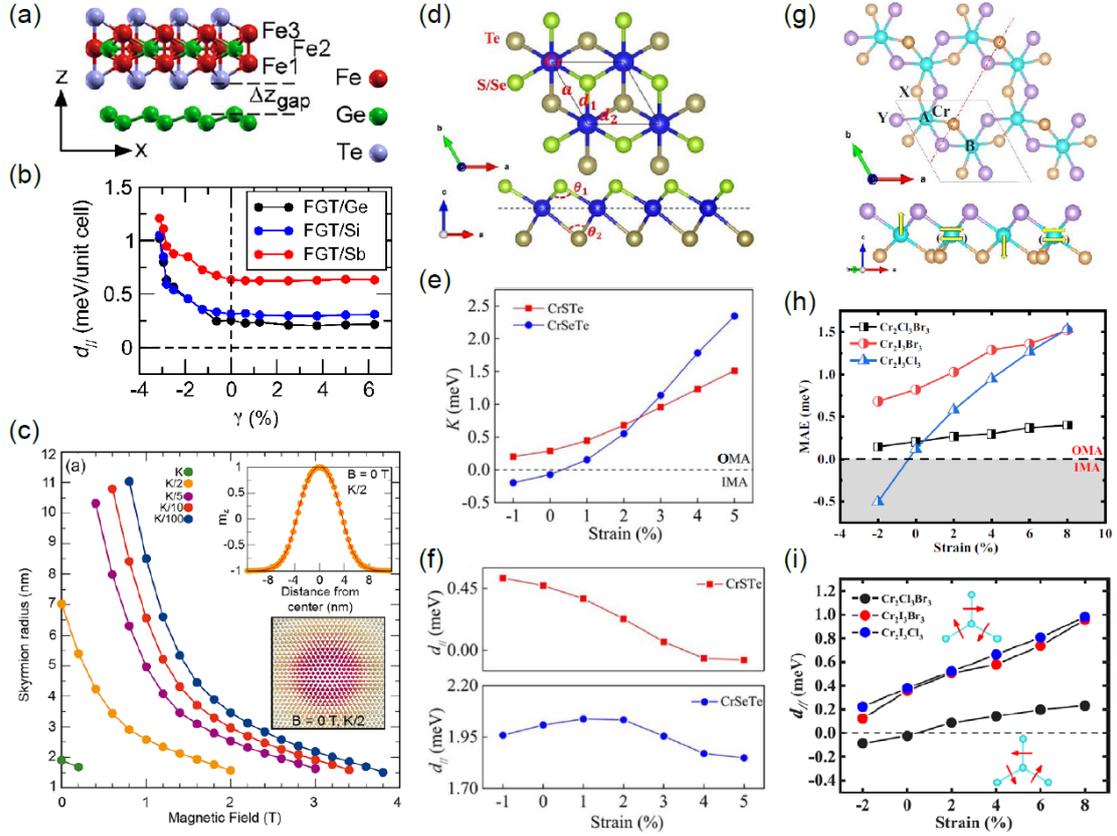

Fig. 5 Strain modulation by calculation. a) Side view of the $Fe_3GeTe_2$/germanene heterostructure. b) The in-plane DMI component ($d_{//}$) as a function of biaxial strain γ. c) Skyrmion radius as a function of magnetic field in strained heterostructure (γ = −3%) with different MAE values, where K = −0.86 meV. The inset presents the skyrmion profile and spin texture at B = 0 T and K/2. Reprinted from Ref. 67. Copyright © 2022, American Chemistry Society. d) Top and side views of Janus CrXTe monolayer. e) The magnetic anisotropy K and (f) $d_{//}$ as a function of strain in Janus CrXTe monolayers. Reprinted from Ref. 68. Copyright © 2020, American Physical Society. g) Top and side views of $Cr_2X_3Y_3$ monolayers. h) MAE defined as the energy difference between in-plane and out-of-plane ferromagnetic states, and i) $d_{//}$ as a function of biaxial strain. The insets present the direction of $d_{//}$ (red arrow). Reprinted from Ref. 69. Copyright © 2022, American Physical Society. OMA: out-of-plane magnetic anisotropy. IMA: in-plane magnetic anisotropy.

5) Stacking and twisting

In 2018, a new mechanism of skyrmions was proposed, that is, the modulation of interlayer magnetic coupling by different atomic locations in ubiquitous moiré pattern [70]. For the 2D vdW materials, the strong covalent bonds ensure the in-plane stability, while the relatively weak vdW forces have the advantage of easy vertical stacking, providing an ideal platform to realize this mechanism[71]. Some recent works attempt 2D bilayer magnets. By twisting, the long-period moiré pattern could be modulated, offering an additional degree of freedom for manipulating magnetism. It was proposed that various skyrmion phases can be stabilized in the twisted bilayer $CrI_3$ and $CrBr_3$[72]. A rich phase diagram with SMTs was uncovered as a function of the twist angle and DMI, as illustrated in Fig. 6(a) and 6(b). Moreover, the recent calculation predicted a multiferroic moiré superlattice with robust magnetoelectric coupling and magnetic bimerons in the 3R-type stacking $LaBr_2$ bilayer with small twisting angles[71]. As illustrated in Fig. 6(c)-6(e), uneven interlayer exchange coupling driven by moiré pattern leads to magnetic topological textures stabilized in domains coupled with staggered electric polarization. By relative twisting or interlayer strain to control interlayer exchange coupling and spatial symmetry, these topological magnetic textures could be tuned effectively[71].

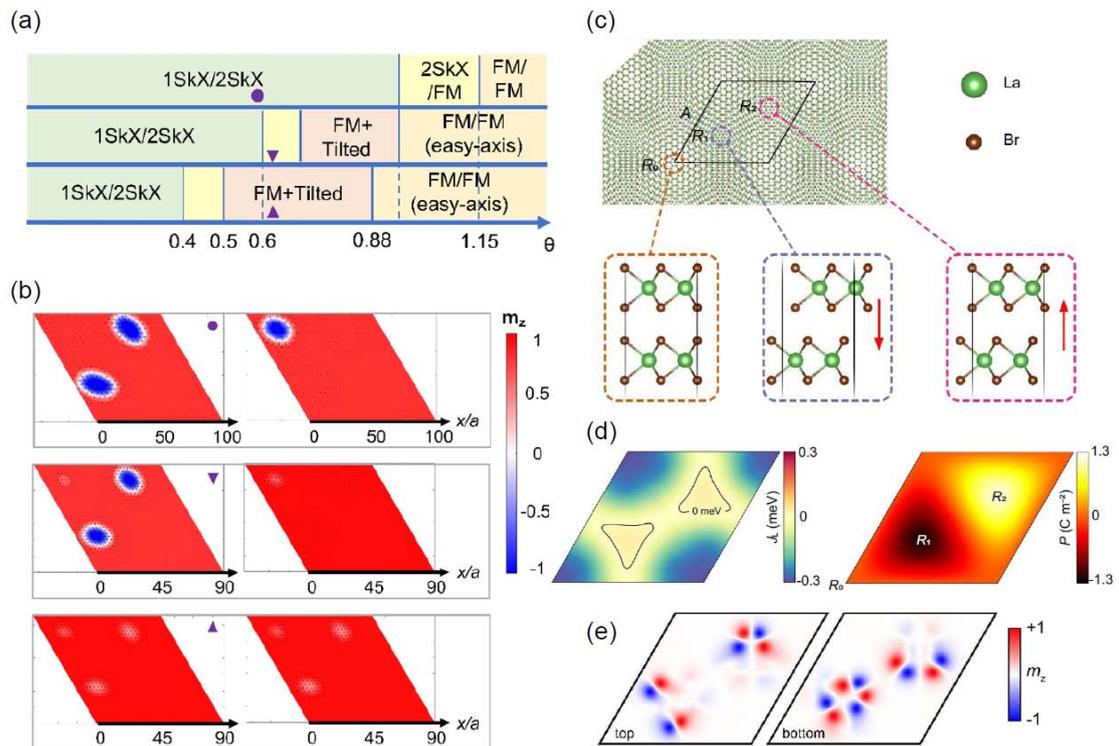

Fig. 6 SMTs in twisted bilayers. a) Phase diagram of twisted bilayer $CrBr_3$ as a function of the twist angle θ, where the three panels from top to bottom correspond to different DMI with the decreasing value. b) From top to bottom: magnetic textures of 1SkX/2SkX, 2SkX/FM, FM+tilted states marked in (a). The two layers are presented on the left and right panels respectively. Reprinted from Ref. 72. Copyright © 2021, American Chemistry Society. c) The moiré superlattice of 3R-type $LaBr_2$ bilayer from twisting. The different interlayer configurations in three local regions ($R_0$, $R_1$, $R_2$) of the moiré superlattice are amplified to show below. d) The interlayer exchange interaction ($J_L$) and electric polarization (P) textures of the corresponding rhombic region in (c). e) The magnetic texture at the twisting angle θ = 0.57°. Reprinted from Ref. 71. Copyright © 2022, The Authors.

## SUMMARY AND PROSPECTIVE

In summary, we review the recent progresses on some representative modulation approaches of SMTs in 2D vdW materials and their heterostructures. In experimental investigations, temperature, magnetic field, sample thickness and electric current have been confirmed to be able to tune and control SMTs effectively. In theoretical calculations, tremendous efforts have been devoted to searching more efficient modulations. Among them, the electric-field control of SMTs in 2D multiferroics is much highlighted. In addition, strain and stacking style are also expected to be efficient approaches due to the unique 2D character of vdW layered materials. All these predictions are still waiting to be verified by experiments. It should be mentioned that besides the aforementioned modulation approaches, there are also other control methods. For example, the formation of skyrmions was proposed in the $CrI_3$ monolayer by substituting I atoms with Cl atoms to break the inversion symmetry[73]. In addition, the various stoichiometric ratios induce different crystal structures with different magnetic sublattices for 2D $Fe_nGeTe_2$, which influence magnetism[74,75]. The skyrmion bubbles have been observed in $Fe_3GeTe_2$ and $Fe_5GeTe_2$[17,76]. It is highly anticipated that more effective modulations could be achieved to control SMTs. The breakthrough on this aspect will greatly help to the technological applications on future spintronics.

On the other hand, from the viewpoints of SMTs' controllable properties, SMTs' creation, annihilation and mobility, as the technological keys for future spintronics applications, have attracted tremendous attention. Moreover, the size of single SMT and the density of SMT lattice have been discussed extensively. There remain another two points worth noting. One is the topological charge, as the fingerprint for topologically untrivial object. A changeable topological charge means breaking topological protection, which is generally hard to be realized. However, a switch between two topologically different stable states is also highly promising for nonvolatile information devices. The attempts on this point have been proposed and even realized in some experiments[64,77,78]. The other is the internal degrees of freedom—vorticity and helicity, which greatly enrich the morphology of SMT. The tunable internal degree of freedom implies the continuous deformation allowed by topological protection, and enables potential functionality by external stimuli with invariant topological property. The calculation predicted that for the frustration-induced SMT in centrosymmetric magnet, its internal degrees of freedom can be manipulated by applying magnetic field and electric field with appropriate orientation and magnitude[79,80]. The very recent experimental observation reported unconventional helicity polarization of skyrmions in vdW $Fe_{5-\delta}GeTe_2$[81], where the skyrmion helicity is controlled by both the dipolar interaction and DMI. The helicity polarization can be attributed to the localized DMI originating from non-uniform regions or deficiencies[81].

**ACKNOWLEDGMENTS**

This work is supported by the Natural Science Foundation of Jiangsu Province (BK20221451), the National Natural Science Foundation of China (11834002, 12274070), and the Fundamental Research Funds for the Central Universities.